\documentclass[aps,11pt,onecolumn,nofootinbib,preprintnumbers, superscriptaddress]{revtex4-2}
\pdfoutput=1

\usepackage[height=8.85in,width=6.45in]{geometry}

\usepackage{tikz}
\usepackage[utf8]{inputenc}
\usepackage[T1]{fontenc}
\usepackage{amsmath}
\usepackage{amssymb}
\usepackage{mathtools}
\usepackage{physics}
\usepackage{enumitem}

\allowdisplaybreaks

\usepackage{slashed}
\usepackage{braket}
\usepackage[usenames,dvipsnames,svgnames,table]{xcolor}
\usepackage[colorlinks,citecolor=DarkGreen,linkcolor=FireBrick,urlcolor=FireBrick,linktocpage,breaklinks=true]{hyperref}
\hypersetup{
  pdfstartview={FitH},    % fits the width of the page to the window
  pdftitle={Thermal spectral function asymptotics and black hole singularity in holography},    % title
  pdfauthor={Hewei Frederic Jia, Mukund Rangamani},     % author
}
\usepackage{cleveref}
\usepackage{graphicx}
\usetikzlibrary{cd,3d}		% commutative diagrams and 3d
\usetikzlibrary{calc}		% Calculations in tikz
\usetikzlibrary{math} %needed tikz library for setting variables
\usetikzlibrary{intersections}
\usetikzlibrary{decorations.pathmorphing}
\usetikzlibrary{decorations.pathreplacing}
\usetikzlibrary{decorations.markings}

\tikzset{snake it/.style={decorate, decoration=snake}}
\tikzset{->-/.style={decoration={
    markings,
    mark=at position .5 with {\arrow{>}}},postaction={decorate}}}
\tikzset{-<-/.style={decoration={
    markings,
    mark=at position .5 with {\arrow{<}}},postaction={decorate}}}

\usepackage[style=base]{subcaption} % incompatible with subfigure and subfig. Latter are known to be broken. Don't use.
\usepackage{float}
\usepackage{afterpage}
\newcommand{\prn}[1]{\left (#1 \right)}
\newcommand{\prnbig}[1]{\Big(#1  \Big)}

\newcommand{\Gret}{K}
\newcommand{\VBcl}{\mathcal{W}}

\newcommand{\cmexp}{\sigma}
\newcommand{\bdy}{\mathrm{bdy}}

\newcommand{\freq}{\mathfrak{w}}
\newcommand{\mom}{\mathfrak{q}}

\newcommand{\lm}{\mathrm{L.M.}}
\newcommand{\rp}{r_{+}}
\newcommand{\ts}{C} %two-sided correlator

\newcommand{\Podd}{P_{\mathrm{odd}}}

\newcommand{\pha}{\vartheta}

\newcommand{\BSX}{\mathcal{X}}

\newcommand{\BSXm}{\hat{\BSX}}

\newcommand{\mc}{\mathcal{C}} %monodromy contour%
\newcommand{\mr}{\zeta} %momentum ratio
\newcommand{\vcl}{v}
\newcommand{\rhonp}{\rho_{\mathrm{np}}} 

\tikzset{
  midarrow/.style={
    postaction={decorate},
    decoration={markings, mark=at position #1 with {\arrow{Stealth}}}
  },
  midarrow/.default=0.5
}

\newcommand{\Half}[1]{\frac{#1}{2}}

\newcommand{\delspl}{\mathrm{D}_{\mathrm{spl}}}

\newcommand{\SAdS}[1]{Schwarzschild-AdS$_{#1}$}
\newcommand{\CFT}[1]{CFT$_{#1}$}

\newcommand{\floor}[1]{\lfloor #1 \rfloor}

\begin{document}

\title{Thermal spectral function asymptotics and black hole singularity in holography}

\author{Hewei Frederic Jia}
\email{heweifred@gmail.com}
\affiliation{Institute for Advanced Study, Tsinghua University, Beijing, 100084, China}

\author{Mukund Rangamani}
\email{mukund@physics.ucdavis.edu}
\affiliation{Center for Quantum Mathematics and Physics (QMAP),
  Department of Physics \& Astronomy, University of California, Davis, CA 95616, USA}

\begin{abstract}
  We investigate the analytic structure of thermal spectral function of holographic CFTs, synthesizing recent developments into a set of observations about its asymptotics.
  Specifically, for a class of scalar primaries with integral dimension, we demonstrate factorization of the exact spectral function into a polynomial piece, which captures the vacuum dynamics, and a non-perturbative
  piece, which controls its asymptotics. Using exact WKB techniques, we derive a transseries expression for the latter. We use this information to deduce the singular loci of a spatially averaged
  thermofield double correlator in the complex time plane. Such singularities have been argued to encode information regarding the black hole singularity in the dual spacetime. Our results give a refinement of these statements by capturing the momentum dependence\@.
\end{abstract}

\maketitle

%%%%%%%%%%%%%%%%%%%%%%%%%%%%%%%%%%%%%%%%%%%%%%%%%%%

%~~~~~~~~~~~~~~~~~~~~~~~~~~~~~~~~~~~~~~~~~~~~~~~
\section{Introduction}\label{sec:intro}
%~~~~~~~~~~~~~~~~~~~~~~~~~~~~~~~~~~~~~~~~~~~~~~

Thermal correlators of a relativistic quantum field theory on Minkowski
spacetime are constrained by Kubo-Martin-Schwinger (KMS) relations. Intuitively, temperature provides an infrared (IR) cut-off, effectively gapping
out most (but not all) of the degrees of freedom. Probing the system at energies much above this IR scale will show deviations from the vacuum result.

Of interest to us are the \emph{non-perturbative} deviations away from this vacuum answer in the context of conformal field theories (CFTs). These corrections turn out to be governed by a transseries expansion, which we determine explicitly for specific holographic CFT correlators.  Our observations build on the work on thermal bootstrap~\cite{Iliesiu:2018fao}, exact analytic approaches to black hole perturbations~\cite{Aminov:2020yma,Bonelli:2022ten,Lisovyy:2022flm,Dodelson:2022yvn}, and rely on exploiting exact WKB techniques~\cite{Iwaki:2014vad,Iwaki:2015xyz}, to determine the asymptotics of the spectral function.

Part of our motivation is to glean information regarding the encoding of the black hole singularity in thermal correlators of holographic CFTs. A key observation was made two decades ago in~\cite{Fidkowski:2003nf} (see also~\cite{Louko:2000tp,Kraus:2002iv}) and was systematized using WKB analysis in~\cite{Festuccia:2005pi,Amado:2008hw,Festuccia:2008zx}. In the past year, several developments have spurred further progress. Firstly, the earlier discussions, which were limited to heavy operators, were extended to light operators using the operator product expansion (OPE)~\cite{Ceplak:2024bja,Afkhami-Jeddi:2025wra,Ceplak:2025dds}. Secondly, analysis of low-dimensional toy models of holography (e.g., the SYK model) allows one to examine the behavior as one deviates from classical general relativity~\cite{Dodelson:2024atp,Dodelson:2025jff}. We also analyze light operators, but work in a different regime of parameters compared to~\cite{Afkhami-Jeddi:2025wra,Dodelson:2025jff}.

Our primary result is a transseries expression for the thermal spectral function. From it, we infer the analytic structure of the (spatially averaged) thermal Green's function in the time domain. Specifically, we find a set of singularities (generalizing the results aforementioned), which appear to be connected with the presence of the black hole singularity in the dual geometry.
We highlight here just the primary features, leaving the details of our analysis for a separate publication~\cite{Jia:2026ryl}.

%~~~~~~~~~~~~~~~~~~~~~~~~~~~~~~~~~~~~~~~~~~~~~~~
\section{Thermal spectral function in CFTs}\label{sec:spectral}
%~~~~~~~~~~~~~~~~~~~~~~~~~~~~~~~~~~~~~~~~~~~~~~

Consider a \CFT{d} at temperature $T$ on $\mathbb{R}^{d-1,1}$ and let $\mathcal{O}_\Delta$ be a scalar conformal primary. We will be interested in the thermal retarded Green's function
$\Gret(t,\vb{x})  \equiv
  -i\, \Theta(t)\,
  \expval{[\mathcal{O}_\Delta(t,\vb{x}),
        \mathcal{O}_\Delta(0,0)]}_\beta$.  For now, we work in frequency space with dimensionless frequency and momenta
\begin{equation}
\freq = \frac{\omega}{2\pi T}\,, \qquad \mom = \frac{\abs{\vb{k}}}{2\pi T}\,.
\end{equation}
The spectral function is obtained from the Fourier transformed Green's function through
\begin{equation}\label{eq:specfn}
\rho(\freq,\mom) = -i \prnbig{\Gret(\freq,\mom) - \Gret(-\freq,\mom)}\,.
\end{equation}

We will be interested in the behavior of the spectral function in a large momentum limit,
\begin{equation}
\lm: \; \freq, \; \mom \to \infty \,,
\qquad
\zeta \equiv \frac{\mom}{\freq} \,, \quad \text{fixed}\,.
\end{equation}
Our analysis will focus on the timelike regime $ 0< \mr <1$, though we will make some remarks about the limiting cases. The case $\mr =0$ has been analyzed recently in~\cite{Afkhami-Jeddi:2025wra} (for holographic \CFT{4}). The lightlike limit $\mr =1$ has also been well-studied~\cite{Festuccia:2008zx} (for quasinormal modes). Our analysis, therefore, straddles the intermediate regime between these extremes and will provide an interpolation between them.

One can deduce the following regarding $\rho(\freq,\mom)$ in the large momentum limit:
\begin{itemize}[wide,left=0pt]
  \item The limit is primarily governed by the vacuum  (i.e., $T=0$) spectral function. Deviations therefrom can be determined from the thermal OPE~\cite{Iliesiu:2018fao} and take the form
  \begin{equation}\label{eq:OPEasym}
  \rho\prn{\freq,\mom} \;
  \overset{\lm}{\sim} \;
  \Theta\prn{\abs{\freq}-\mom} \;
  (\freq^2 - \mom^2)^{\Delta -\frac{d}{2}}\;
  \prnbig{1 + \order{\mom^{-\alpha}}} + \order{e^{-\mom}} \,.
  \end{equation}
  The subleading power-law corrections are controlled by the OPE and have support on the timelike regime $\mr \in[0,1)$. In contrast, the exponentially small corrections are not completely controlled by the OPE; they
  can also have support in the spacelike regime  (cf.~\cite{Manenti:2019wxs}).\footnote{For developments of analytic thermal bootstrap, exploiting the analytic structure of thermal correlators and for applications in the holographic context see~\cite{Alday:2020eua,Buric:2025anb,Barrat:2025nvu,Buric:2025fye,Barrat:2025twb}.}
  \item For holographic \CFT{d} with even $d$, and operators with the special integer dimensions~\cite{Caron-Huot:2009ypo,Dodelson:2023vrw}
  \begin{equation}\label{eq:integer-special-dimensions}
  \Delta \equiv \frac{d}{2} + \nu  \in \Big\{\Half{d}, \Half{d}+1, \cdots, d-1, d \Big\} \equiv \delspl,
  \end{equation}
  there are only exponentially small corrections to the leading vacuum spectral density, i.e., no $\mom^{-\alpha}$ corrections indicated in~\eqref{eq:OPEasym}.\footnote{ While we are discussing scalar operators here, it is also worth noting that $\Delta =d$ is related to the transverse traceless spin-2 component of the stress tensor in the holographic context.\label{fn:stspin2}}
\end{itemize}
For a holographic \CFT{4} with $\Delta \in \delspl$ defined in~\eqref{eq:integer-special-dimensions}, we observe a factorization of the exact spectral function into a vacuum and a piece $\rhonp$ that captures non-perturbative corrections (cf.~\eqref{eq:factorconj})
\begin{equation}\label{eq:rhofactor}
\rho(\freq,\mom) = (\freq^2 - \mom^2)^{\Delta -\frac{d}{2}}\ \rhonp(\freq,\mom).
\end{equation}

\noindent
\textbf{Claim:} The asymptotics of $\rhonp$ for $\Delta \in \delspl$ is obtained as a transseries
\begin{equation}\label{eq:sfasym}
\rhonp(\freq,\mr\,\freq)
\overset{\freq \to \infty}{\sim}
1 + \sum_{r=1}^{\infty}
\sum_{s\overset{2}{\,=\,}-r}^{\infty}
e^{(-r\pi-s\,\vcl)\freq}
\sum^{r}_{q\overset{2}{\,=\,}r-2\,
\floor{\frac{r}{2}}}
\Re \prn{e^{i q(\pi-\vcl) \freq} P_{rsq}(\freq)}.
\end{equation}
The parameter controlling asymptotics is $\vcl$ which is related to $\mr$ via
\begin{equation}\label{eq:vclzeta}
\vcl =
\frac{\sqrt{2\pi}\, \Gamma\prn{\frac{5}{4}}}{\Gamma\prn{\frac{7}{4}}} \;
\mr^{\Half{3}} \;
{}_2 F_1\prn{\frac{1}{4},\frac{3}{4};\frac{7}{4};\mr^2} \,.
\end{equation}
This parameter smoothly interpolates between $0$ and $\pi$ as $\mr$ varies from the zero momentum limit to the lightlike limit. We note that~\eqref{eq:sfasym} is the analog of eq.(23) in~\cite{Afkhami-Jeddi:2025wra} in the case of taking both $\freq,\mom \to \infty$ with their ratio fixed, instead of taking $\freq \to \infty$ at fixed $\mom$. Compared to that case, we are finding perturbative expansions of integer instead of fractional powers in the non-perturbative sector. We first justify Eqs.~\eqref{eq:rhofactor} and~\eqref{eq:sfasym} and then discuss what they imply vis-\`a-vis the signatures of the black hole singularity. In~\cite{Jia:2026ryl} we first motive the factorization using the thermal OPE analysis, and then derive the transseries by exploiting an exact expression for the thermal two-point function obtained in~\cite{Jia:2024zes}. 

%~~~~~~~~~~~~~~~~~~~~~~~~~~~~~~~~~~~~~~~~~~~~~~~
\section{Holographic correlators from Virasoro block}\label{sec:holvir}
%~~~~~~~~~~~~~~~~~~~~~~~~~~~~~~~~~~~~~~~~~~~~~~

Our starting point is an expression of the retarded Green's function derived using the Virasoro block~\cite{Bonelli:2022ten,Lisovyy:2022flm} in the form presented in~\cite{Jia:2024zes}
\begin{equation}\label{eq:KwqVir}
\Gret(\freq,\mom)
=
\frac{2}{\nu} \, \frac{\Gamma\prn{-\nu}}{\Gamma\prn{\nu}}\,
\prod_{\eta = \pm 1}\,
\frac{\Gamma\prn{\frac{1}{2} \,(1+\nu + i\,\freq) + \eta\, \cmexp}}{
  \Gamma\prn{\frac{1}{2} \,(1-\nu-i\,\freq) + \eta\, \cmexp}} \, e^{\VBcl_{\bdy}}  \,.
\end{equation}
This result relies on mapping the holographic computation to the BPZ equation  for a level-$2$ null vector decoupling in an auxiliary two-dimensional CFT~\cite{Bonelli:2022ten,Lisovyy:2022flm}. The physical parameters $\freq$ and $\nu$ appear as conformal weights of external operators. $\VBcl_{\bdy}(\freq,\mom) = 2\, \partial_\nu \VBcl$ is the derivative of the semiclassical Virasoro block. $\cmexp(\freq,\mom)$ is also determined from $\VBcl$ using the Zamolodchikov relation. However, as we argue below, it has a geometric meaning as well. It is related to the trace of the monodromy matrix between the horizon and the AdS boundary, a fact that enables us exploit WKB techniques to compute it.

The result holds for any holographic \CFT{d}, but for $d=4$ the blocks involved are 4-point blocks
\begin{equation}\label{eq:Wsblock-qnm}
\VBcl(x) = \begin{tikzpicture}[scale= .35, baseline = .5ex]
  \coordinate (inf) at (-2,-2);
  \coordinate (one) at (-2,2);
  \coordinate (inf-one) at (0,0);
  \coordinate (xr-zero) at (4,0);
  \coordinate (midint) at (2,0);
  \coordinate (xr) at (6,2);
  \coordinate (zero) at (6,-2);
  \draw (inf) -- (inf-one);
  \draw  (one) -- (inf-one);
  \draw (inf-one) -- (xr-zero);
  \draw (xr-zero) -- (xr);
  \draw (xr-zero) -- (zero);
  \node [below left] at (inf) {\scriptsize $0 \,
      (u=0)$};
  \node[below left] at (-1.8,-3.1)  {\scriptsize{singularity}} ;
  \node [above left] at (one) {\scriptsize $\frac{\freq}{2} \, (u=1)$};
  \node [above left] at (-1,3.3) {\scriptsize {complex horizon}};
  \node [below] at (midint) {\scriptsize $\cmexp(\freq,\mom)$};
  \node [above right] at (xr) {\scriptsize $\nu \, (u=2)$};
  \node [above right] at (5.7,3.2) {\scriptsize boundary};
  \node [below right] at (zero) {\scriptsize $\frac{i\freq}{2} (u=\infty)$};
  \node [below right] at (7,-3.3) {\scriptsize horizon};
\end{tikzpicture}
\end{equation}
The parameter $\cmexp$ determines the dimension of the intermediate operator exchanged in this auxiliary correlator. It can be fixed in terms of the semiclassical block, so one can, in principle, use series expansion of the Virasoro block to work out a series expansion for $\cmexp$ in the cross-ratio parameter. However, in most cases of interest, the cross-ratio itself is fixed by the observable (and the holographic dual geometry). The strategy works best when there is a hierarchy of geometric scales (small black holes, near-extremality, etc.).  The reader can find examples in the aforementioned references.

From~\eqref{eq:KwqVir} we can deduce that the spectral function~\eqref{eq:specfn} can be expressed for $\nu \in \mathbb{Z}_{\geq 0}$ as
\begin{equation}\label{eq:holspfn}
\rho(\freq,\mom)
=
\mathsf{N}(\nu)\, P_{2\nu}(\freq,\cmexp)\, e^{\VBcl_{\bdy}}\,
\times
\frac{\sinh(\pi\freq)}{\cosh(\pi\freq)+ (-1)^{ \nu}\cos(2\pi\cmexp)}
\,,
\qquad
\mathsf{N}(\nu) \equiv \frac{2}{\nu} \, \frac{\Gamma\prn{-\nu}}{\Gamma\prn{\nu}}\,,
\end{equation}
where $P_{2\nu}(\freq,\cmexp)$ is a degree $2\nu$ polynomial.
For $\Delta \in \delspl$~\eqref{eq:integer-special-dimensions}, we have been able to show that $P_{2\nu}(\freq,\cmexp)\, e^{\VBcl_{\bdy}} \sim (\freq^2 - \mom^2)^\nu$ in $d=4$.\footnote{These claims are verified in~\cite{Jia:2026ryl} using the semiclassical Virasoro blocks obtained by Zamolodchikov recursion. They imply interesting identities for the block at cross-ratio $x= \frac{1}{2}$.}
In addition, the result holds for $\nu =0$ in any $d$ with $P_{2\nu}(\freq,\cmexp)\, e^{\VBcl_{\bdy}} =1$. Finally, a familiar case is $d=2$ with $\Delta=1,2$, where
$P_{2\nu}(\freq,\cmexp)\, e^{\VBcl_{\bdy}} =(\freq^2 - \mom^2)^{\nu}$ and $\cmexp = i\mom/2$.\footnote{A similar simplification occurs for R-current correlators of $\mathcal{N} =4$ SYM at vanishing momentum $\mom =0$~\cite{Myers:2007we}, a fact that was exploited recently in~\cite{Afkhami-Jeddi:2025wra} to motivate the general picture. Likewise, as noted in~\cref{fn:stspin2} the case $\nu =2$ in $d=4$ gives the correlator for transverse traceless spin-2 polarization of the stress tensor, for these components are described holographically by a massless Klein-Gordon scalar.}

For general $d$ we conjecture that the spectral function for $\Delta \in \delspl$ factorizes into a vacuum contribution and a non-perturbative piece $\rhonp$ with
\begin{equation}\label{eq:factorconj}
\rho(\freq,\mom) =
\mathsf{N}(\nu)\, (\freq^2 - \mom^2)^{\Delta -\frac{d}{2}}\ \rhonp(\freq,\mom)
\,,
\qquad
\rhonp(\freq,\mom) \coloneqq \frac{\sinh(\pi\freq)}{\cosh(\pi\freq)+ (-1)^{ \nu}\cos(2\pi\cmexp)}\,.
\end{equation}
We will turn to determining the asymptotics of $\rhonp$ defined in~\eqref{eq:factorconj}. The $\cmexp$ dependence contained in $\cos(2\pi\cmexp)$ is proportional to the trace of the monodromy matrix around the two singular points at the horizon and at infinity:\footnote{The connection problem between the singular points at the horizon and the asymptotic boundary, for instance, determines the retarded Green's function of the dual CFT $K(\freq, \mom)$~\cite{Son:2002sd}.}
\begin{equation}\label{eq:Mhorbdy}
\Tr(M_{\mathrm{bdy},\mathrm{hor}}) =  -2\cos(2\pi\cmexp) \in \mathbb{R}\,.
\end{equation}
The monodromy invariant can be analyzed using exact WKB. Doing so we obtain an exact transseries expansion for it, as we explain below.

%~~~~~~~~~~~~~~~~~~~~~~~~~~~~~~~~~~~~~~~~~~~~~~~
\section{Exact WKB of black hole wave equations}\label{sec:wkb}
%~~~~~~~~~~~~~~~~~~~~~~~~~~~~~~~~~~~~~~~~~~~~~~

\begin{figure}[htbp!]
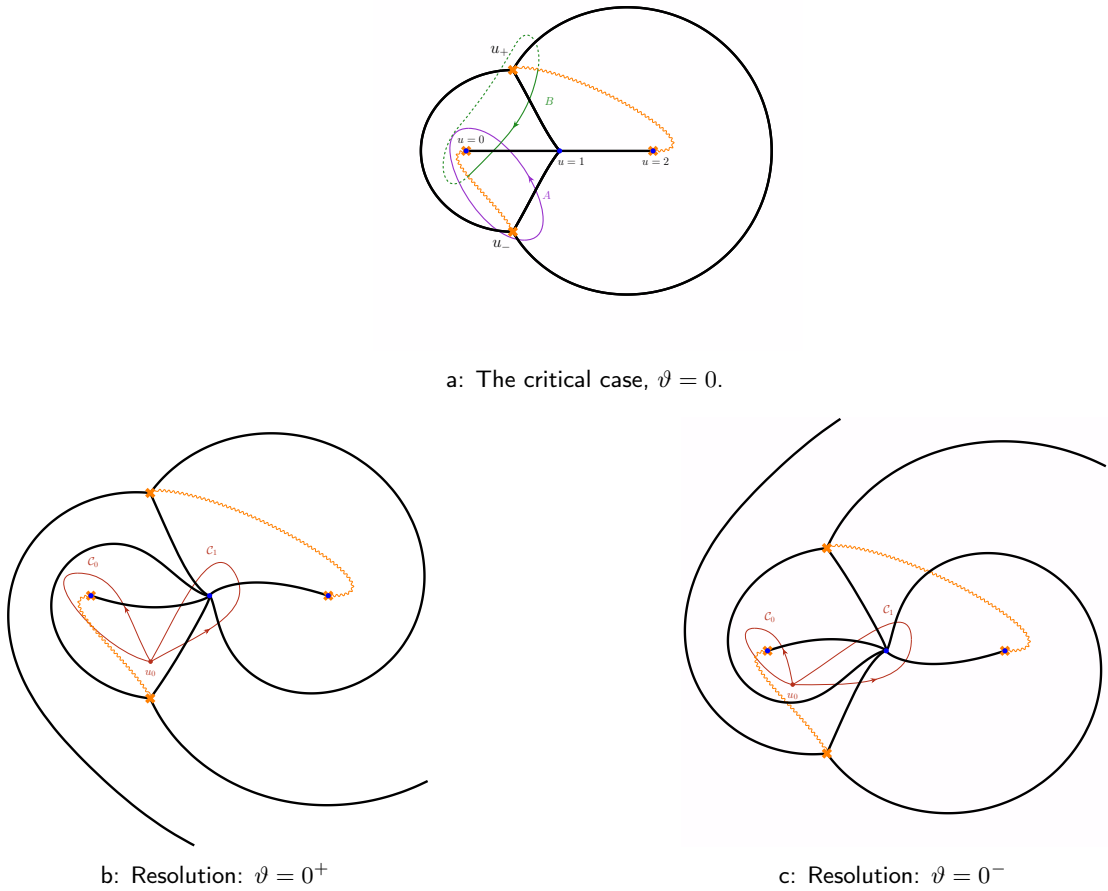

  \centering
  \begin{subfigure}[b]{0.5\textwidth}
    \centering
    \scalebox{.3}{\input{timelike-crit-tikz}}
    \caption{The critical case, $\pha  = 0$.}
  \end{subfigure}

  \vspace{0.5em}

  \begin{subfigure}[b]{0.4\textwidth}
    \centering
    \scalebox{.3}{\input{timelike-pos-tikz}}
    \caption{Resolution: $\pha = 0^{+}$}
  \end{subfigure}
  \hfill
  \begin{subfigure}[b]{0.5\textwidth}
    \centering
    \scalebox{.3}{\input{timelike-neg-tikz}}
    \caption{Resolution: $\pha = 0^{-}$}
  \end{subfigure}
  \caption{The Stokes graphs for~\eqref{eq:Qpot} in the timelike regime $0<\zeta <1$ with $\pha = \arg(\freq) = \arg(\mom)$. The thick black lines are Stokes lines that demarcate the different Stokes regions on the $u$ plane. The turning points, singularities, placement of branch cuts, monodromy contours $\mathcal{C}_{0,1}$ centered at base point $u_0$, and the A and B cycles are indicated for reference. }
  \label{fig:Stokes-graphs}
\end{figure}

Consider a minimally coupled scalar field of mass $m^2 = \Delta \,(\Delta -4)$ in the planar-\SAdS{5} background. The line element is
\begin{equation}
ds^{2} =
-r^{2}\, f(r) \,dt^{2} + \frac{dr^{2}}{r^{2}f(r)} + r^{2}\,  d\vb{x}^2_{d-1}
\,, \qquad
f(r) = 1 - \prn{\frac{\rp}{r}}^{4} \,.
\end{equation}
The black hole is dual to the thermal state at $T = \frac{\rp}{\pi}$ and the scalar field is dual to a conformal primary of dimension $\Delta$. Working with coordinates
\begin{equation}
u =  \frac{2\, r^2}{r^2 - r_+^2}\,,
\end{equation}
the Klein-Gordon equation can be presented in Schr\"odinger form as
\begin{equation}\label{eq:SchrAds5}
\psi''(u) - \freq^2\, Q(u) \, \psi(u) = 0\,,
\qquad
Q(u) = Q_0(u) + \freq^{-2}\, Q_2(u)\,.
\end{equation}
We have exploited the fact that we wish to work in the limit $\freq \to \infty$. The potentials are
\begin{equation}\label{eq:Qpot}
Q_{0}(u) =
\frac{u^2 - 4\,(u-1)\,\mr^2}{4\,(2-u)\,(u-1)^2 \,u}\,,
\qquad
Q_{2}(u) =
-\frac{(u^2-2\,u+2)^2}{4\,u^2\,(u-1)^2\,(u-2)^2} +
\frac{\nu^2}{4\,(u-2)^2\,(u-1)}\,.
\end{equation}
This form is amenable to exact WKB analysis with turning points at
$u_\pm =  2 \,\mr^2 \pm 2\,i\,\mr \,\sqrt{1-\mr^2}$. In addition, we have simple poles at the black hole singularity ($u=0$) and a complex horizon ($u=1$ or $r = i\,\rp$), respectively, and double poles at the horizon ($u=\infty$) and the timelike boundary ($u=2$), cf.~\eqref{eq:Wsblock-qnm}. We will take a different perspective on this differential equation from what was employed in~\cref{sec:holvir}. In particular, it is clear from~\eqref{eq:Wsblock-qnm} that the monodromy required in~\eqref{eq:Mhorbdy} can equivalently be computed by examining the connection problem between the complex horizon $u=1$ and the curvature singularity $u=0$:
$M_{\mathrm{bdy}, \mathrm{hor}} = M_{u=0}\, M_{u=1}$. This is our first inkling that the asymptotics of $\rho$ could have information about the black hole singularity.

To compute the monodromy itself, we exploit the procedure explained in~\cite{Kawai:2005aas}.\footnote{A review of exact WKB techniques, specifically adapted for computing the monodromy asymptotics is described in~\cite[Section 4]{Jia:2026ryl}.} Suppose we want to compute the monodromy around a singular point $u_{i}$ along a monodromy contour $\mc_{i}$ with base point $u_{0}$, cf.~\cref{fig:Stokes-graphs}. We work with the WKB solutions normalized at $u_{0}$
\begin{equation}
\psi_{\pm}(u,\freq) = \frac{1}{\sqrt{\Podd(u,\freq)}}
\exp\prn{\pm \int^{u}_{u_{0}} du^{\prime} \Podd(u',\freq)} \,,
\qquad
\Podd(u,\freq) = \sum_{\mathrm{odd}\; n \geq -1}\, P_n(u) \,\freq^{-n}\,.
\end{equation}
$\Podd$ is the odd part of the all-orders `momentum', with  $P_{-1} =\sqrt{Q_0}$ being the classical momentum used in standard WKB. The monodromy contour of interest in general will cross Stokes lines emanating from some branch point $v_\alpha$ (the simple poles and turning points) at $\{u^{(\alpha)}\}$.
The connection matrix at the crossing can be computed using WKB solutions normalized at $v_\alpha$ and accounting for the change of reference point. Taking the product over all such crossings of Stokes lines we obtain the required monodromy matrix. This will depend on the WKB periods owing to the changes in the reference point in the above procedure.

For the problem at hand, the case of real momentum corresponds to a critical phase of the WKB analysis.  One can address the latter issue by resolving the critical phase.\footnote{We find that one has to utilize the median symbols introduced in~\cite{Delabaere:1997srq} so that the Borel resummed series satisfy the same reality conditions that the asymptotic series do. The median symbols are computed after giving a small phase to the WKB parameter and then taking a suitable limit by averaging over the sign of this phase.} The resolved Stokes graphs\footnote{It is also worth noting that the topology of the Stokes graphs changes between timelike and spacelike regimes, clearly distinguishing the two.} shown in~\cref{fig:Stokes-graphs} involve crossing of the monodromy contour across the Stokes lines joining the turning points to the singularity, complex horizon, and  the horizon. All told, we determine an exact answer for $\cos(2\pi\cmexp)$ in terms of the median Borel-resummed Voros symbol $\BSXm_{A}$
\begin{equation}\label{eq:cmexp-asymptotics}
\cos(2\pi\cmexp) = \abs{\BSXm_{A}}^2 e^{-\pi\freq}\cos(\pi\nu) + \sqrt{\prn{1+\abs{\BSXm_{A}}^2 e^{-2\pi\freq}} \prn{1 + \abs{\BSXm_{A}}^{-2}}} \Re\prn{\BSXm_{A} e^{i \pi \freq}} \,.
\end{equation}

The Voros symbol is the exponential of the all orders WKB period $X_{A} = \exp\prn{\oint_A \, \Podd(u)\,du}$~\cite{Voros:1983ret}. It may be viewed as the non-perturbative action computed to all-orders, and may require Borel resummmation of the perturbative series.\footnote{Our notation is as follows: $X_{A}$ is formal asymptotic power series from WKB analysis, and $\BSX_{A}$ is its Borel summed counterpart. The median Borel sum is indicated with a hat decoration. } 
The cycle ($A$) encircles the turning point $u_-$ and the singularity, cf.~\cref{fig:Stokes-graphs}. The asymptotics of the median Borel-resummed Voros symbol $\BSXm$ recovers the original Voros symbol $X$. The parameter $\vcl$ introduced earlier in~\eqref{eq:vclzeta} is the classical WKB period around this cycle. Subleading corrections depend on the higher periods. Finding the asymptotics of~\eqref{eq:cmexp-asymptotics} and then using~\eqref{eq:factorconj} leads to the result quoted in~\eqref{eq:sfasym}. Details of this derivation,
both computing the period integrals, and assembling the Voros symbol can be found in~\cite[Section 5]{Jia:2026ryl}.
Specifically, the leading term of~\eqref{eq:sfasym} reads
\begin{equation}\label{eq:rhoser1}
\rhonp(\freq,\mr \,\freq)
\overset{\freq \to \infty}{\sim}
1 + 2\, (-1)^{\nu+1} \, e^{-(\pi-\vcl)\freq} \, \Re(e^{i(\pi-\vcl)\freq}P(\freq)) + \cdots \,,
\end{equation}
with $P(\freq)$ determined by higher periods. The leading perturbative correction in this non-perturbative sector is found to be
\begin{equation}\label{eq:Pomega}
P(\freq) =
\exp\prn{
  e^{i\,\frac{\pi}{4}}\,
  \frac{4\,\sqrt{\pi}\, \Gamma\prn{\frac{5}{4}}}{\Gamma\prn{\frac{3}{4}}}\,\frac{2\,\nu^2 + \mr^{-2} -2}{16\, \freq} \,
  \mr^{-\frac{1}{2}}\,
  \bqty{
    (1-\mr^2)^{-\frac{1}{4}} - (1-\mr^2)^{-\frac{3}{4}}
  }
  + \order{\freq^{-3}}
}\,.
\end{equation}
%

%~~~~~~~~~~~~~~~~~~~~~~~~~~~~~~~~~~~~~~~~~~~~~~~
\section{Inklings of the black hole singularity?}\label{sec:bhsing}
%~~~~~~~~~~~~~~~~~~~~~~~~~~~~~~~~~~~~~~~~~~~~~~

With the asymptotics of the spectral function at hand, let us turn to the second question motivating our analysis, viz., the signatures of the black hole singularity.  In~\cite{Fidkowski:2003nf} it was observed that radial spacelike geodesics connecting the two asymptotic boundaries anchored at fixed boundary time are repelled by the singularity. They are bounded by a pair of limiting null geodesics.  The presence of this distinguished pair, which is referred to as the \emph{bouncing null geodesic},\footnote{Strictly speaking, the null geodesics do not bounce,  but rather terminate at the singularity. On the other hand the spacelike geodesics do bounce off before reaching the singularity. } is a clear geometric signature of the Lorentzian geometry. Its presence implies a critical real time $t_c = \frac{\beta}{2}\, \cot\prn{\frac{\pi}{d}}$ beyond which no \emph{real spacelike geodesic} connects the two asymptotic boundaries.

This geodesic signature indicates that the thermofield (or two-sided) correlator $C_\Delta(t,\vb{x}) = \expval{\mathcal{O}_\Delta(t + i\frac{\beta}{2}, \vb{x})\, \mathcal{O}_\Delta(t, \vb{x})}_\beta$ has a divergence $(t-t_c)^{-2\,\Delta}$ on a secondary sheet for heavy operators~\cite{Fidkowski:2003nf}. These observations were revisited and sharpened in~\cite{Festuccia:2005pi} (see also~\cite{Amado:2008hw}) who argued that the signature is more directly visible in Fourier domain should one examine the behavior as $\freq \to -i\infty$.

These statements generally have been argued to require $\Delta  \gg 1$, with~\cite{Amado:2008hw} being an exception. In the latter work it is argued that the eikonal approximation can be used or any $\Delta$ to identify the singularities of the correlator in the complex time domain.
Recently, it was argued in~\cite{Ceplak:2024bja} that the stress tensor exchange in the thermal OPE decomposition (at fixed spatial separation) encodes a striking feature of $t_c$ even for $\Delta \sim \order{1}$. Likewise, at fixed spatial momentum~\cite{Afkhami-Jeddi:2025wra} and~\cite{Dodelson:2025jff} argue for a similar signature of the bouncing geodesic. The former uses WKB to demonstrate exponentially small corrections to the vacuum answer (extending old observations~\cite{Teaney:2006nc,Caron-Huot:2009ypo}), while the latter uses a combination of thermal product formula~\cite{Dodelson:2023vrw}
and asymptotics of quasinormal modes.
These statements can be distilled into the following observation~\cite{Festuccia:2005pi}: The holographic thermal correlator is meromorphic, with a line of poles reaching out along some ray in the complex frequency plane. The direction along which this ray approaches $\abs{\freq} \to \infty$ determines $t_c$.

Our results, which use both WKB and the Virasoro block approach to thermal correlators, extend these statements obtained from $\mr =0$ to the range $0 < \mr < 1$. The thermofield correlator is directly related to the spectral density.\footnote{ In Fourier domain one has the relation
  $\ts(\omega,k) = \frac{\rho(\omega,k)}{2\sinh\prn{\frac{\beta\omega}{2}}}$. }
Using~\eqref{eq:sfasym} we can obtain an analogous transseries  for $\ts(\freq,\mr\, \freq)$. A termwise Fourier transform $\ts_\mr(t) =
  \int_0^\infty\, d\freq\, \cos(\freq\, t)\, \ts(\freq, \mr\, \freq)$
informs us of the singularity structure in the time domain.\footnote{
  The observable $\ts_\mr(t)$ is spatially averaged. Splitting $\vb{x} $ into radial and angular components, we have
  $\ts_\mr(t) = \int_0^\infty\, d\abs{\vb{x}}\, \ts(t + \zeta\, \abs{\vb{x}}, \abs{\vb{x}}, \Omega_{d-2}) $.}
In general, $\ts(t,\vb{x})$ at fixed $\vb{x}$ is guaranteed to be analytic within the strip $\Im(t) \in (-\frac{\beta}{2}, \frac{\beta}{2})$, but has OPE singularities at the edge of the strip. We can analytically continue it as a periodic function to the full complex t-plane using the KMS condition at non-zero $\vb{x}$~\cite{Iliesiu:2018fao}. However, finite momentum or spatial averaging can introduce additional singularities outside the strip.

For $\mr= 0$ the singularities of $\ts_0(t)$, which is at fixed spatial momentum,  are at~\cite{Dodelson:2025jff} (see also~\cite{Ceplak:2024bja})
\begin{equation}\label{eq:trq}
t_{rq} = \frac{i\beta}{2} + r \frac{i\beta}{2} + q\, \frac{\beta}{2}\cot\prn{\frac{\pi}{d}}, \quad
r \in \mathbb{Z}_{\geq 1}, \quad q \in \{-r,-r+2,\dots, r\}\,.
\end{equation}
This gives $t_c = \frac{\beta}{2}\, \cot\prn{\frac{\pi}{d}}$.
For $\mr\neq 0$, our result for $\ts(\freq,\mr\,\freq)$ implies that
$\ts_\mr(t)$ in $d=4$ has singularities at $t = \pm t_{rsq}$, located at
\begin{equation}\label{eq:trsqzeta}
\begin{split}
 & t_{rsq} =
\frac{i \beta}{2} + \prn{r + s\,\frac{v}{\pi}} \frac{i\beta}{2} + q \,\frac{\beta}{2}\prn{1-\frac{v}{\pi}},
\\
 & r
\in \mathbb{Z}_{\geq1}, \quad
s\in -r + 2\, \mathbb{Z}_{\geq 0}, \quad
q \in \{-r,-r+2,\dots, r\}\,.
\end{split}
\end{equation}

Clearly,~\eqref{eq:trsqzeta} agrees with~\eqref{eq:trq} for $d=4$ at $\mr = \vcl = 0$. While the singularities in~\eqref{eq:trq} were argued to lie on codimension-1 loci $\Im(t) = n\,\frac{\beta}{2}$, we see that this is no longer the case for $\mr \neq 0$. In fact, our result is similar to the charged black hole case (probed by a neutral operator) analyzed in~\cite{Dodelson:2025jff,Ceplak:2025dds}, even though we are studying a neutral black hole. While the causal structure of the two black holes is quite different, there is a physical reason why this could have been anticipated.

In the absence of spatial momentum (or for $\mr =0$) one is analyzing radial null geodesics as
in~\cite{Fidkowski:2003nf}. However, when the spatial momentum is activated there is an effective `centrifugal barrier' causing the geodesics to either crash into the singularity, or bounce off a different locus, depending on their conserved energy. This feature is similar to the geodesic motion in the charged black hole case, where the presence of the charge causes the spacelike geodesics to be repelled by the Cauchy horizon (cf.~\cite{Ceplak:2025dds} for a recent discussion). In fact, examining the near-singularity behavior of the effective potential for geodesic motion, charged black hole geodesics with vanishing momentum are akin to neutral black hole geodesics with non-vanishing momentum and vice versa.

Finally, as we approach the lightlike limit, $\mr \to 1$,  we see that the singularities disappear. We believe this is because the limit corresponds to ballistic UV propagation with very highly damped decay probability, so the WKB paths or geodesics stay well away from the singularity.\footnote{ The picture is transparent for global AdS black holes where the geodesics are circular orbits close to the boundary in the large angular momentum limit. In either case, in this limit we have very weakly damped quasinormal modes~\cite{Festuccia:2008zx} for this reason.} Relatedly, the Stokes graph changes topology: the spacetime boundary becomes a turning point for WKB.

%~~~~~~~~~~~~~~~~~~~~~~~~~~~~~~~~~~~~~~~~~~~~~~~
\section{Discussion}\label{sec:discussion}
%~~~~~~~~~~~~~~~~~~~~~~~~~~~~~~~~~~~~~~~~~~~~~~

Our main result is a determination of an exact transseries expression for the thermal spectral function~\eqref{eq:sfasym} using exact WKB methods.
For this we exploited a factorized form of the spectral function for $\Delta \in \delspl$ in holographic CFTs.
It would be interesting to investigate the generality of this factorized form (both for $\Delta \notin \delspl$ and for general CFTs).

Given these asymptotics, we have a prediction for a lattice of singularities on the complex time plane outside the fundamental strip $\Im(t) \in (-\frac{\beta}{2}, \frac{\beta}{2})$ for $\ts_\mr(t)$. The presence of these singularities in the real-time correlator has been linked to the black hole singularity in the geometry. However, it is worth pointing out that the connection needs to be sharpened. The discussions of~\cite{Fidkowski:2003nf,Festuccia:2005pi} make clear the connection with a real Lorentzian section geodesic for $\Delta \gg1$. However, the statements for $\Delta \sim \order{1}$ either derived using the OPE at fixed spatial separation~\cite{Ceplak:2024bja,Ceplak:2025dds}, or derived using asymptotics at fixed spatial momentum as here~\cite{Afkhami-Jeddi:2025wra,Dodelson:2025jff}, rely only on having information about a timescale, $t_c$, extracted from complex WKB trajectories. While the latter do involve eikonal propagation and multiple bounces off the singularity (cf.~\cite{Afkhami-Jeddi:2025wra}),
a more direct connection with the geometry would be desirable. One tantalizing piece of evidence in favor is that the monodromy around the horizon and the boundary is the same as that between the singularity and all the other singular points of the wave equation.\footnote{ For $d>4$ (and for non-scalar probes in $d=4$) the wave equations are Fuchsian with four or more regular singular points. The monodromy around horizon and the boundary can instead be traded for that between the singular point at the black hole singularity and the rest.   }

%\newpage
%~~~~~~~~~~~~~~~~~~~~~~~~~~~~~~~~~~~~~~~~~~~~~~
\begin{acknowledgments}
  It is a pleasure to thank Veronika Hubeny and Zhenbin Yang for illuminating discussions. In addition, we would like to thank Simon Caron-Huot for a conversation that inspired this investigation and helpful comments on a preliminary draft of this manuscript. We also thank Matthew Dodelson, Andrei Parnachev, and Steve Shenker for comments on the draft.
  M.R.~would like to acknowledge support from the Simons Center for Geometry and Physics, Stony Brook University, during the initial stages of this project. M.R.~is also grateful to the
  Aspen Center for Physics, which is supported by National Science Foundation grant PHY-2210452, where some of this research was carried out.
  H.F.J.~acknowledges support from the Tsinghua Shuimu Scholar program.
  M.R.~was supported by U.S. Department of Energy grant DE-SC0009999 and by funds from the University of California.
\end{acknowledgments}

%~~~~~~~~~~~~~~~~~~~~~~~~~~~~~~~~~~~~~~~~~~~~~~
% \vfill
% \hbox{}

% \bibliographystyle{JHEP}
% \bibliography{wkb-refs}

\begin{thebibliography}{34}%
\makeatletter
\providecommand \@ifxundefined [1]{%
 \@ifx{#1\undefined}
}%
\providecommand \@ifnum [1]{%
 \ifnum #1\expandafter \@firstoftwo
 \else \expandafter \@secondoftwo
 \fi
}%
\providecommand \@ifx [1]{%
 \ifx #1\expandafter \@firstoftwo
 \else \expandafter \@secondoftwo
 \fi
}%
\providecommand \natexlab [1]{#1}%
\providecommand \enquote  [1]{``#1''}%
\providecommand \bibnamefont  [1]{#1}%
\providecommand \bibfnamefont [1]{#1}%
\providecommand \citenamefont [1]{#1}%
\providecommand \href@noop [0]{\@secondoftwo}%
\providecommand \href [0]{\begingroup \@sanitize@url \@href}%
\providecommand \@href[1]{\@@startlink{#1}\@@href}%
\providecommand \@@href[1]{\endgroup#1\@@endlink}%
\providecommand \@sanitize@url [0]{\catcode `\\12\catcode `\$12\catcode `\&12\catcode `\#12\catcode `\^12\catcode `\_12\catcode `\%12\relax}%
\providecommand \@@startlink[1]{}%
\providecommand \@@endlink[0]{}%
\providecommand \url  [0]{\begingroup\@sanitize@url \@url }%
\providecommand \@url [1]{\endgroup\@href {#1}{\urlprefix }}%
\providecommand \urlprefix  [0]{URL }%
\providecommand \Eprint [0]{\href }%
\providecommand \doibase [0]{https://doi.org/}%
\providecommand \selectlanguage [0]{\@gobble}%
\providecommand \bibinfo  [0]{\@secondoftwo}%
\providecommand \bibfield  [0]{\@secondoftwo}%
\providecommand \translation [1]{[#1]}%
\providecommand \BibitemOpen [0]{}%
\providecommand \bibitemStop [0]{}%
\providecommand \bibitemNoStop [0]{.\EOS\space}%
\providecommand \EOS [0]{\spacefactor3000\relax}%
\providecommand \BibitemShut  [1]{\csname bibitem#1\endcsname}%
\let\auto@bib@innerbib\@empty
%</preamble>
\bibitem [{\citenamefont {Iliesiu}\ \emph {et~al.}(2018)\citenamefont {Iliesiu}, \citenamefont {Kolo{\u{g}}lu}, \citenamefont {Mahajan}, \citenamefont {Perlmutter},\ and\ \citenamefont {Simmons-Duffin}}]{Iliesiu:2018fao}%
  \BibitemOpen
  \bibfield  {author} {\bibinfo {author} {\bibfnamefont {L.}~\bibnamefont {Iliesiu}}, \bibinfo {author} {\bibfnamefont {M.}~\bibnamefont {Kolo{\u{g}}lu}}, \bibinfo {author} {\bibfnamefont {R.}~\bibnamefont {Mahajan}}, \bibinfo {author} {\bibfnamefont {E.}~\bibnamefont {Perlmutter}},\ and\ \bibinfo {author} {\bibfnamefont {D.}~\bibnamefont {Simmons-Duffin}},\ }\bibfield  {title} {\bibinfo {title} {{The Conformal Bootstrap at Finite Temperature}},\ }\href {https://doi.org/10.1007/JHEP10(2018)070} {\bibfield  {journal} {\bibinfo  {journal} {JHEP}\ }\textbf {\bibinfo {volume} {10}},\ \bibinfo {pages} {070}},\ \Eprint {https://arxiv.org/abs/1802.10266} {arXiv:1802.10266 [hep-th]} \BibitemShut {NoStop}%
\bibitem [{\citenamefont {Aminov}\ \emph {et~al.}(2022)\citenamefont {Aminov}, \citenamefont {Grassi},\ and\ \citenamefont {Hatsuda}}]{Aminov:2020yma}%
  \BibitemOpen
  \bibfield  {author} {\bibinfo {author} {\bibfnamefont {G.}~\bibnamefont {Aminov}}, \bibinfo {author} {\bibfnamefont {A.}~\bibnamefont {Grassi}},\ and\ \bibinfo {author} {\bibfnamefont {Y.}~\bibnamefont {Hatsuda}},\ }\bibfield  {title} {\bibinfo {title} {{Black Hole Quasinormal Modes and Seiberg\textendash{}Witten Theory}},\ }\href {https://doi.org/10.1007/s00023-021-01137-x} {\bibfield  {journal} {\bibinfo  {journal} {Annales Henri Poincare}\ }\textbf {\bibinfo {volume} {23}},\ \bibinfo {pages} {1951} (\bibinfo {year} {2022})},\ \Eprint {https://arxiv.org/abs/2006.06111} {arXiv:2006.06111 [hep-th]} \BibitemShut {NoStop}%
\bibitem [{\citenamefont {Bonelli}\ \emph {et~al.}(2023)\citenamefont {Bonelli}, \citenamefont {Iossa}, \citenamefont {Panea~Lichtig},\ and\ \citenamefont {Tanzini}}]{Bonelli:2022ten}%
  \BibitemOpen
  \bibfield  {author} {\bibinfo {author} {\bibfnamefont {G.}~\bibnamefont {Bonelli}}, \bibinfo {author} {\bibfnamefont {C.}~\bibnamefont {Iossa}}, \bibinfo {author} {\bibfnamefont {D.}~\bibnamefont {Panea~Lichtig}},\ and\ \bibinfo {author} {\bibfnamefont {A.}~\bibnamefont {Tanzini}},\ }\bibfield  {title} {\bibinfo {title} {{Irregular Liouville Correlators and Connection Formulae for Heun Functions}},\ }\href {https://doi.org/10.1007/s00220-022-04497-5} {\bibfield  {journal} {\bibinfo  {journal} {Commun. Math. Phys.}\ }\textbf {\bibinfo {volume} {397}},\ \bibinfo {pages} {635} (\bibinfo {year} {2023})},\ \Eprint {https://arxiv.org/abs/2201.04491} {arXiv:2201.04491 [hep-th]} \BibitemShut {NoStop}%
\bibitem [{\citenamefont {Lisovyy}\ and\ \citenamefont {Naidiuk}(2022)}]{Lisovyy:2022flm}%
  \BibitemOpen
  \bibfield  {author} {\bibinfo {author} {\bibfnamefont {O.}~\bibnamefont {Lisovyy}}\ and\ \bibinfo {author} {\bibfnamefont {A.}~\bibnamefont {Naidiuk}},\ }\bibfield  {title} {\bibinfo {title} {{Perturbative connection formulas for Heun equations}},\ }\href {https://doi.org/10.1088/1751-8121/ac9ba7} {\bibfield  {journal} {\bibinfo  {journal} {J. Phys. A}\ }\textbf {\bibinfo {volume} {55}},\ \bibinfo {pages} {434005} (\bibinfo {year} {2022})},\ \Eprint {https://arxiv.org/abs/2208.01604} {arXiv:2208.01604 [math-ph]} \BibitemShut {NoStop}%
\bibitem [{\citenamefont {Dodelson}\ \emph {et~al.}(2022)\citenamefont {Dodelson}, \citenamefont {Grassi}, \citenamefont {Iossa}, \citenamefont {Panea~Lichtig},\ and\ \citenamefont {Zhiboedov}}]{Dodelson:2022yvn}%
  \BibitemOpen
  \bibfield  {author} {\bibinfo {author} {\bibfnamefont {M.}~\bibnamefont {Dodelson}}, \bibinfo {author} {\bibfnamefont {A.}~\bibnamefont {Grassi}}, \bibinfo {author} {\bibfnamefont {C.}~\bibnamefont {Iossa}}, \bibinfo {author} {\bibfnamefont {D.}~\bibnamefont {Panea~Lichtig}},\ and\ \bibinfo {author} {\bibfnamefont {A.}~\bibnamefont {Zhiboedov}},\ }\bibfield  {title} {\bibinfo {title} {{Holographic thermal correlators from supersymmetric instantons}},\ }\href@noop {} {\  (\bibinfo {year} {2022})},\ \Eprint {https://arxiv.org/abs/2206.07720} {arXiv:2206.07720 [hep-th]} \BibitemShut {NoStop}%
\bibitem [{\citenamefont {Iwaki}\ and\ \citenamefont {Nakanishi}(2014)}]{Iwaki:2014vad}%
  \BibitemOpen
  \bibfield  {author} {\bibinfo {author} {\bibfnamefont {K.}~\bibnamefont {Iwaki}}\ and\ \bibinfo {author} {\bibfnamefont {T.}~\bibnamefont {Nakanishi}},\ }\bibfield  {title} {\bibinfo {title} {{Exact WKB analysis and cluster algebras}},\ }\href {https://doi.org/10.1088/1751-8113/47/47/474009} {\bibfield  {journal} {\bibinfo  {journal} {J. Phys. A}\ }\textbf {\bibinfo {volume} {47}},\ \bibinfo {pages} {474009} (\bibinfo {year} {2014})},\ \Eprint {https://arxiv.org/abs/1401.7094} {arXiv:1401.7094 [math.CA]} \BibitemShut {NoStop}%
\bibitem [{\citenamefont {Iwaki}\ and\ \citenamefont {Nakanishi}(2015)}]{Iwaki:2015xyz}%
  \BibitemOpen
  \bibfield  {author} {\bibinfo {author} {\bibfnamefont {K.}~\bibnamefont {Iwaki}}\ and\ \bibinfo {author} {\bibfnamefont {T.}~\bibnamefont {Nakanishi}},\ }\bibfield  {title} {\bibinfo {title} {Exact wkb analysis and cluster algebras ii: Simple poles, orbifold points, and generalized cluster algebras},\ }\href {https://doi.org/10.1093/imrn/rnv270} {\bibfield  {journal} {\bibinfo  {journal} {International Mathematics Research Notices}\ }\textbf {\bibinfo {volume} {2016}},\ \bibinfo {pages} {4375–4417} (\bibinfo {year} {2015})}\BibitemShut {NoStop}%
\bibitem [{\citenamefont {Fidkowski}\ \emph {et~al.}(2004)\citenamefont {Fidkowski}, \citenamefont {Hubeny}, \citenamefont {Kleban},\ and\ \citenamefont {Shenker}}]{Fidkowski:2003nf}%
  \BibitemOpen
  \bibfield  {author} {\bibinfo {author} {\bibfnamefont {L.}~\bibnamefont {Fidkowski}}, \bibinfo {author} {\bibfnamefont {V.}~\bibnamefont {Hubeny}}, \bibinfo {author} {\bibfnamefont {M.}~\bibnamefont {Kleban}},\ and\ \bibinfo {author} {\bibfnamefont {S.}~\bibnamefont {Shenker}},\ }\bibfield  {title} {\bibinfo {title} {{The Black hole singularity in AdS / CFT}},\ }\href {https://doi.org/10.1088/1126-6708/2004/02/014} {\bibfield  {journal} {\bibinfo  {journal} {JHEP}\ }\textbf {\bibinfo {volume} {02}},\ \bibinfo {pages} {014}},\ \Eprint {https://arxiv.org/abs/hep-th/0306170} {arXiv:hep-th/0306170} \BibitemShut {NoStop}%
\bibitem [{\citenamefont {Louko}\ \emph {et~al.}(2000)\citenamefont {Louko}, \citenamefont {Marolf},\ and\ \citenamefont {Ross}}]{Louko:2000tp}%
  \BibitemOpen
  \bibfield  {author} {\bibinfo {author} {\bibfnamefont {J.}~\bibnamefont {Louko}}, \bibinfo {author} {\bibfnamefont {D.}~\bibnamefont {Marolf}},\ and\ \bibinfo {author} {\bibfnamefont {S.~F.}\ \bibnamefont {Ross}},\ }\bibfield  {title} {\bibinfo {title} {{On geodesic propagators and black hole holography}},\ }\href {https://doi.org/10.1103/PhysRevD.62.044041} {\bibfield  {journal} {\bibinfo  {journal} {Phys. Rev. D}\ }\textbf {\bibinfo {volume} {62}},\ \bibinfo {pages} {044041} (\bibinfo {year} {2000})},\ \Eprint {https://arxiv.org/abs/hep-th/0002111} {arXiv:hep-th/0002111} \BibitemShut {NoStop}%
\bibitem [{\citenamefont {Kraus}\ \emph {et~al.}(2003)\citenamefont {Kraus}, \citenamefont {Ooguri},\ and\ \citenamefont {Shenker}}]{Kraus:2002iv}%
  \BibitemOpen
  \bibfield  {author} {\bibinfo {author} {\bibfnamefont {P.}~\bibnamefont {Kraus}}, \bibinfo {author} {\bibfnamefont {H.}~\bibnamefont {Ooguri}},\ and\ \bibinfo {author} {\bibfnamefont {S.}~\bibnamefont {Shenker}},\ }\bibfield  {title} {\bibinfo {title} {{Inside the horizon with AdS / CFT}},\ }\href {https://doi.org/10.1103/PhysRevD.67.124022} {\bibfield  {journal} {\bibinfo  {journal} {Phys. Rev. D}\ }\textbf {\bibinfo {volume} {67}},\ \bibinfo {pages} {124022} (\bibinfo {year} {2003})},\ \Eprint {https://arxiv.org/abs/hep-th/0212277} {arXiv:hep-th/0212277} \BibitemShut {NoStop}%
\bibitem [{\citenamefont {Festuccia}\ and\ \citenamefont {Liu}(2006)}]{Festuccia:2005pi}%
  \BibitemOpen
  \bibfield  {author} {\bibinfo {author} {\bibfnamefont {G.}~\bibnamefont {Festuccia}}\ and\ \bibinfo {author} {\bibfnamefont {H.}~\bibnamefont {Liu}},\ }\bibfield  {title} {\bibinfo {title} {{Excursions beyond the horizon: Black hole singularities in Yang-Mills theories. I.}},\ }\href {https://doi.org/10.1088/1126-6708/2006/04/044} {\bibfield  {journal} {\bibinfo  {journal} {JHEP}\ }\textbf {\bibinfo {volume} {04}},\ \bibinfo {pages} {044}},\ \Eprint {https://arxiv.org/abs/hep-th/0506202} {arXiv:hep-th/0506202} \BibitemShut {NoStop}%
\bibitem [{\citenamefont {Amado}\ and\ \citenamefont {Hoyos-Badajoz}(2008)}]{Amado:2008hw}%
  \BibitemOpen
  \bibfield  {author} {\bibinfo {author} {\bibfnamefont {I.}~\bibnamefont {Amado}}\ and\ \bibinfo {author} {\bibfnamefont {C.}~\bibnamefont {Hoyos-Badajoz}},\ }\bibfield  {title} {\bibinfo {title} {{AdS black holes as reflecting cavities}},\ }\href {https://doi.org/10.1088/1126-6708/2008/09/118} {\bibfield  {journal} {\bibinfo  {journal} {JHEP}\ }\textbf {\bibinfo {volume} {09}},\ \bibinfo {pages} {118}},\ \Eprint {https://arxiv.org/abs/0807.2337} {arXiv:0807.2337 [hep-th]} \BibitemShut {NoStop}%
\bibitem [{\citenamefont {Festuccia}\ and\ \citenamefont {Liu}(2009)}]{Festuccia:2008zx}%
  \BibitemOpen
  \bibfield  {author} {\bibinfo {author} {\bibfnamefont {G.}~\bibnamefont {Festuccia}}\ and\ \bibinfo {author} {\bibfnamefont {H.}~\bibnamefont {Liu}},\ }\bibfield  {title} {\bibinfo {title} {{A Bohr-Sommerfeld quantization formula for quasinormal frequencies of AdS black holes}},\ }\href {https://doi.org/10.1166/asl.2009.1029} {\bibfield  {journal} {\bibinfo  {journal} {Adv. Sci. Lett.}\ }\textbf {\bibinfo {volume} {2}},\ \bibinfo {pages} {221} (\bibinfo {year} {2009})},\ \Eprint {https://arxiv.org/abs/0811.1033} {arXiv:0811.1033 [gr-qc]} \BibitemShut {NoStop}%
\bibitem [{\citenamefont {{\v{C}}eplak}\ \emph {et~al.}(2024)\citenamefont {{\v{C}}eplak}, \citenamefont {Liu}, \citenamefont {Parnachev},\ and\ \citenamefont {Valach}}]{Ceplak:2024bja}%
  \BibitemOpen
  \bibfield  {author} {\bibinfo {author} {\bibfnamefont {N.}~\bibnamefont {{\v{C}}eplak}}, \bibinfo {author} {\bibfnamefont {H.}~\bibnamefont {Liu}}, \bibinfo {author} {\bibfnamefont {A.}~\bibnamefont {Parnachev}},\ and\ \bibinfo {author} {\bibfnamefont {S.}~\bibnamefont {Valach}},\ }\bibfield  {title} {\bibinfo {title} {{Black hole singularity from OPE}},\ }\href {https://doi.org/10.1007/JHEP10(2024)105} {\bibfield  {journal} {\bibinfo  {journal} {JHEP}\ }\textbf {\bibinfo {volume} {10}},\ \bibinfo {pages} {105}},\ \Eprint {https://arxiv.org/abs/2404.17286} {arXiv:2404.17286 [hep-th]} \BibitemShut {NoStop}%
\bibitem [{\citenamefont {Afkhami-Jeddi}\ \emph {et~al.}(2025)\citenamefont {Afkhami-Jeddi}, \citenamefont {Caron-Huot}, \citenamefont {Chakravarty},\ and\ \citenamefont {Maloney}}]{Afkhami-Jeddi:2025wra}%
  \BibitemOpen
  \bibfield  {author} {\bibinfo {author} {\bibfnamefont {N.}~\bibnamefont {Afkhami-Jeddi}}, \bibinfo {author} {\bibfnamefont {S.}~\bibnamefont {Caron-Huot}}, \bibinfo {author} {\bibfnamefont {J.}~\bibnamefont {Chakravarty}},\ and\ \bibinfo {author} {\bibfnamefont {A.}~\bibnamefont {Maloney}},\ }\bibfield  {title} {\bibinfo {title} {{Imprint of the black hole singularity on thermal two-point functions}},\ }\href@noop {} {\  (\bibinfo {year} {2025})},\ \Eprint {https://arxiv.org/abs/2510.21673} {arXiv:2510.21673 [hep-th]} \BibitemShut {NoStop}%
\bibitem [{\citenamefont {{\v{C}}eplak}\ \emph {et~al.}(2025)\citenamefont {{\v{C}}eplak}, \citenamefont {Liu}, \citenamefont {Parnachev},\ and\ \citenamefont {Valach}}]{Ceplak:2025dds}%
  \BibitemOpen
  \bibfield  {author} {\bibinfo {author} {\bibfnamefont {N.}~\bibnamefont {{\v{C}}eplak}}, \bibinfo {author} {\bibfnamefont {H.}~\bibnamefont {Liu}}, \bibinfo {author} {\bibfnamefont {A.}~\bibnamefont {Parnachev}},\ and\ \bibinfo {author} {\bibfnamefont {S.}~\bibnamefont {Valach}},\ }\bibfield  {title} {\bibinfo {title} {{Fooling the Censor: Going beyond inner horizons with the OPE}},\ }\href@noop {} {\  (\bibinfo {year} {2025})},\ \Eprint {https://arxiv.org/abs/2511.09638} {arXiv:2511.09638 [hep-th]} \BibitemShut {NoStop}%
\bibitem [{\citenamefont {Dodelson}(2025)}]{Dodelson:2024atp}%
  \BibitemOpen
  \bibfield  {author} {\bibinfo {author} {\bibfnamefont {M.}~\bibnamefont {Dodelson}},\ }\bibfield  {title} {\bibinfo {title} {{Ringdown in the SYK model}},\ }\href {https://doi.org/10.21468/SciPostPhys.19.3.081} {\bibfield  {journal} {\bibinfo  {journal} {SciPost Phys.}\ }\textbf {\bibinfo {volume} {19}},\ \bibinfo {pages} {081} (\bibinfo {year} {2025})},\ \Eprint {https://arxiv.org/abs/2408.05790} {arXiv:2408.05790 [hep-th]} \BibitemShut {NoStop}%
\bibitem [{\citenamefont {Dodelson}\ \emph {et~al.}(2025)\citenamefont {Dodelson}, \citenamefont {Iossa},\ and\ \citenamefont {Karlsson}}]{Dodelson:2025jff}%
  \BibitemOpen
  \bibfield  {author} {\bibinfo {author} {\bibfnamefont {M.}~\bibnamefont {Dodelson}}, \bibinfo {author} {\bibfnamefont {C.}~\bibnamefont {Iossa}},\ and\ \bibinfo {author} {\bibfnamefont {R.}~\bibnamefont {Karlsson}},\ }\bibfield  {title} {\bibinfo {title} {{Bouncing off a stringy singularity}},\ }\href@noop {} {\  (\bibinfo {year} {2025})},\ \Eprint {https://arxiv.org/abs/2511.09616} {arXiv:2511.09616 [hep-th]} \BibitemShut {NoStop}%
\bibitem [{\citenamefont {Jia}\ and\ \citenamefont {Rangamani}(2026)}]{Jia:2026ryl}%
  \BibitemOpen
  \bibfield  {author} {\bibinfo {author} {\bibfnamefont {H.~F.}\ \bibnamefont {Jia}}\ and\ \bibinfo {author} {\bibfnamefont {M.}~\bibnamefont {Rangamani}},\ }\bibfield  {title} {\bibinfo {title} {{Exact holographic thermal spectral functions: OPE, non-perturbative corrections, and  black hole singularity}},\ }\href@noop {} {\  (\bibinfo {year} {2026})},\ \Eprint {https://arxiv.org/abs/2604.10803} {2604.10803 [hep-th]} \BibitemShut {NoStop}%
\bibitem [{\citenamefont {Manenti}(2020)}]{Manenti:2019wxs}%
  \BibitemOpen
  \bibfield  {author} {\bibinfo {author} {\bibfnamefont {A.}~\bibnamefont {Manenti}},\ }\bibfield  {title} {\bibinfo {title} {{Thermal CFTs in momentum space}},\ }\href {https://doi.org/10.1007/JHEP01(2020)009} {\bibfield  {journal} {\bibinfo  {journal} {JHEP}\ }\textbf {\bibinfo {volume} {01}},\ \bibinfo {pages} {009}},\ \Eprint {https://arxiv.org/abs/1905.01355} {arXiv:1905.01355 [hep-th]} \BibitemShut {NoStop}%
\bibitem [{\citenamefont {Alday}\ \emph {et~al.}(2021)\citenamefont {Alday}, \citenamefont {Kologlu},\ and\ \citenamefont {Zhiboedov}}]{Alday:2020eua}%
  \BibitemOpen
  \bibfield  {author} {\bibinfo {author} {\bibfnamefont {L.~F.}\ \bibnamefont {Alday}}, \bibinfo {author} {\bibfnamefont {M.}~\bibnamefont {Kologlu}},\ and\ \bibinfo {author} {\bibfnamefont {A.}~\bibnamefont {Zhiboedov}},\ }\bibfield  {title} {\bibinfo {title} {{Holographic correlators at finite temperature}},\ }\href {https://doi.org/10.1007/JHEP06(2021)082} {\bibfield  {journal} {\bibinfo  {journal} {JHEP}\ }\textbf {\bibinfo {volume} {06}},\ \bibinfo {pages} {082}},\ \Eprint {https://arxiv.org/abs/2009.10062} {arXiv:2009.10062 [hep-th]} \BibitemShut {NoStop}%
\bibitem [{\citenamefont {Buri{\'c}}\ \emph {et~al.}(2025{\natexlab{a}})\citenamefont {Buri{\'c}}, \citenamefont {Gusev},\ and\ \citenamefont {Parnachev}}]{Buric:2025anb}%
  \BibitemOpen
  \bibfield  {author} {\bibinfo {author} {\bibfnamefont {I.}~\bibnamefont {Buri{\'c}}}, \bibinfo {author} {\bibfnamefont {I.}~\bibnamefont {Gusev}},\ and\ \bibinfo {author} {\bibfnamefont {A.}~\bibnamefont {Parnachev}},\ }\bibfield  {title} {\bibinfo {title} {{Thermal holographic correlators and KMS condition}},\ }\href {https://doi.org/10.1007/JHEP09(2025)053} {\bibfield  {journal} {\bibinfo  {journal} {JHEP}\ }\textbf {\bibinfo {volume} {09}},\ \bibinfo {pages} {053}},\ \Eprint {https://arxiv.org/abs/2505.10277} {arXiv:2505.10277 [hep-th]} \BibitemShut {NoStop}%
\bibitem [{\citenamefont {Barrat}\ \emph {et~al.}(2025{\natexlab{a}})\citenamefont {Barrat}, \citenamefont {Bozkurt}, \citenamefont {Marchetto}, \citenamefont {Miscioscia},\ and\ \citenamefont {Pomoni}}]{Barrat:2025nvu}%
  \BibitemOpen
  \bibfield  {author} {\bibinfo {author} {\bibfnamefont {J.}~\bibnamefont {Barrat}}, \bibinfo {author} {\bibfnamefont {D.~N.}\ \bibnamefont {Bozkurt}}, \bibinfo {author} {\bibfnamefont {E.}~\bibnamefont {Marchetto}}, \bibinfo {author} {\bibfnamefont {A.}~\bibnamefont {Miscioscia}},\ and\ \bibinfo {author} {\bibfnamefont {E.}~\bibnamefont {Pomoni}},\ }\bibfield  {title} {\bibinfo {title} {{The analytic bootstrap at finite temperature}},\ }\href@noop {} {\  (\bibinfo {year} {2025}{\natexlab{a}})},\ \Eprint {https://arxiv.org/abs/2506.06422} {arXiv:2506.06422 [hep-th]} \BibitemShut {NoStop}%
\bibitem [{\citenamefont {Buri{\'c}}\ \emph {et~al.}(2025{\natexlab{b}})\citenamefont {Buri{\'c}}, \citenamefont {Gusev},\ and\ \citenamefont {Parnachev}}]{Buric:2025fye}%
  \BibitemOpen
  \bibfield  {author} {\bibinfo {author} {\bibfnamefont {I.}~\bibnamefont {Buri{\'c}}}, \bibinfo {author} {\bibfnamefont {I.}~\bibnamefont {Gusev}},\ and\ \bibinfo {author} {\bibfnamefont {A.}~\bibnamefont {Parnachev}},\ }\bibfield  {title} {\bibinfo {title} {{Holographic Correlators from Thermal Bootstrap}},\ }\href@noop {} {\  (\bibinfo {year} {2025}{\natexlab{b}})},\ \Eprint {https://arxiv.org/abs/2508.08373} {arXiv:2508.08373 [hep-th]} \BibitemShut {NoStop}%
\bibitem [{\citenamefont {Barrat}\ \emph {et~al.}(2025{\natexlab{b}})\citenamefont {Barrat}, \citenamefont {Bozkurt}, \citenamefont {Marchetto}, \citenamefont {Miscioscia},\ and\ \citenamefont {Pomoni}}]{Barrat:2025twb}%
  \BibitemOpen
  \bibfield  {author} {\bibinfo {author} {\bibfnamefont {J.}~\bibnamefont {Barrat}}, \bibinfo {author} {\bibfnamefont {D.~N.}\ \bibnamefont {Bozkurt}}, \bibinfo {author} {\bibfnamefont {E.}~\bibnamefont {Marchetto}}, \bibinfo {author} {\bibfnamefont {A.}~\bibnamefont {Miscioscia}},\ and\ \bibinfo {author} {\bibfnamefont {E.}~\bibnamefont {Pomoni}},\ }\bibfield  {title} {\bibinfo {title} {{Analytic thermal bootstrap meets holography}},\ }\href@noop {} {\  (\bibinfo {year} {2025}{\natexlab{b}})},\ \Eprint {https://arxiv.org/abs/2510.20894} {arXiv:2510.20894 [hep-th]} \BibitemShut {NoStop}%
\bibitem [{\citenamefont {Caron-Huot}(2009)}]{Caron-Huot:2009ypo}%
  \BibitemOpen
  \bibfield  {author} {\bibinfo {author} {\bibfnamefont {S.}~\bibnamefont {Caron-Huot}},\ }\bibfield  {title} {\bibinfo {title} {{Asymptotics of thermal spectral functions}},\ }\href {https://doi.org/10.1103/PhysRevD.79.125009} {\bibfield  {journal} {\bibinfo  {journal} {Phys. Rev. D}\ }\textbf {\bibinfo {volume} {79}},\ \bibinfo {pages} {125009} (\bibinfo {year} {2009})},\ \Eprint {https://arxiv.org/abs/0903.3958} {arXiv:0903.3958 [hep-ph]} \BibitemShut {NoStop}%
\bibitem [{\citenamefont {Dodelson}\ \emph {et~al.}(2024)\citenamefont {Dodelson}, \citenamefont {Iossa}, \citenamefont {Karlsson},\ and\ \citenamefont {Zhiboedov}}]{Dodelson:2023vrw}%
  \BibitemOpen
  \bibfield  {author} {\bibinfo {author} {\bibfnamefont {M.}~\bibnamefont {Dodelson}}, \bibinfo {author} {\bibfnamefont {C.}~\bibnamefont {Iossa}}, \bibinfo {author} {\bibfnamefont {R.}~\bibnamefont {Karlsson}},\ and\ \bibinfo {author} {\bibfnamefont {A.}~\bibnamefont {Zhiboedov}},\ }\bibfield  {title} {\bibinfo {title} {{A thermal product formula}},\ }\href {https://doi.org/10.1007/JHEP01(2024)036} {\bibfield  {journal} {\bibinfo  {journal} {JHEP}\ }\textbf {\bibinfo {volume} {01}},\ \bibinfo {pages} {036}},\ \Eprint {https://arxiv.org/abs/2304.12339} {arXiv:2304.12339 [hep-th]} \BibitemShut {NoStop}%
\bibitem [{\citenamefont {Jia}\ and\ \citenamefont {Rangamani}(2024)}]{Jia:2024zes}%
  \BibitemOpen
  \bibfield  {author} {\bibinfo {author} {\bibfnamefont {H.~F.}\ \bibnamefont {Jia}}\ and\ \bibinfo {author} {\bibfnamefont {M.}~\bibnamefont {Rangamani}},\ }\bibfield  {title} {\bibinfo {title} {{Holographic thermal correlators and quasinormal modes from semiclassical Virasoro blocks}},\ }\href {https://doi.org/10.1007/JHEP12(2024)047} {\bibfield  {journal} {\bibinfo  {journal} {JHEP}\ }\textbf {\bibinfo {volume} {12}},\ \bibinfo {pages} {047}},\ \Eprint {https://arxiv.org/abs/2408.05208} {arXiv:2408.05208 [hep-th]} \BibitemShut {NoStop}%
\bibitem [{\citenamefont {Myers}\ \emph {et~al.}(2007)\citenamefont {Myers}, \citenamefont {Starinets},\ and\ \citenamefont {Thomson}}]{Myers:2007we}%
  \BibitemOpen
  \bibfield  {author} {\bibinfo {author} {\bibfnamefont {R.~C.}\ \bibnamefont {Myers}}, \bibinfo {author} {\bibfnamefont {A.~O.}\ \bibnamefont {Starinets}},\ and\ \bibinfo {author} {\bibfnamefont {R.~M.}\ \bibnamefont {Thomson}},\ }\bibfield  {title} {\bibinfo {title} {{Holographic spectral functions and diffusion constants for fundamental matter}},\ }\href {https://doi.org/10.1088/1126-6708/2007/11/091} {\bibfield  {journal} {\bibinfo  {journal} {JHEP}\ }\textbf {\bibinfo {volume} {11}},\ \bibinfo {pages} {091}},\ \Eprint {https://arxiv.org/abs/0706.0162} {arXiv:0706.0162 [hep-th]} \BibitemShut {NoStop}%
\bibitem [{\citenamefont {Son}\ and\ \citenamefont {Starinets}(2002)}]{Son:2002sd}%
  \BibitemOpen
  \bibfield  {author} {\bibinfo {author} {\bibfnamefont {D.~T.}\ \bibnamefont {Son}}\ and\ \bibinfo {author} {\bibfnamefont {A.~O.}\ \bibnamefont {Starinets}},\ }\bibfield  {title} {\bibinfo {title} {{Minkowski space correlators in AdS / CFT correspondence: Recipe and applications}},\ }\href {https://doi.org/10.1088/1126-6708/2002/09/042} {\bibfield  {journal} {\bibinfo  {journal} {JHEP}\ }\textbf {\bibinfo {volume} {09}},\ \bibinfo {pages} {042}},\ \Eprint {https://arxiv.org/abs/hep-th/0205051} {arXiv:hep-th/0205051} \BibitemShut {NoStop}%
\bibitem [{\citenamefont {Kawai}\ and\ \citenamefont {Takei}(2005)}]{Kawai:2005aas}%
  \BibitemOpen
  \bibfield  {author} {\bibinfo {author} {\bibfnamefont {T.}~\bibnamefont {Kawai}}\ and\ \bibinfo {author} {\bibfnamefont {Y.}~\bibnamefont {Takei}},\ }\href@noop {} {\emph {\bibinfo {title} {Algebraic Analysis of Singular Perturbation Theory}}},\ \bibinfo {series} {Translations of Mathematical Monographs}, Vol.\ \bibinfo {volume} {227}\ (\bibinfo  {publisher} {American Mathematical Society},\ \bibinfo {address} {Providence, RI},\ \bibinfo {year} {2005})\ p.\ \bibinfo {pages} {129},\ \bibinfo {note} {iwanami Series in Modern Mathematics}\BibitemShut {NoStop}%
\bibitem [{\citenamefont {Delabaere}\ \emph {et~al.}(1997)\citenamefont {Delabaere}, \citenamefont {Dillinger},\ and\ \citenamefont {Pham}}]{Delabaere:1997srq}%
  \BibitemOpen
  \bibfield  {author} {\bibinfo {author} {\bibfnamefont {E.}~\bibnamefont {Delabaere}}, \bibinfo {author} {\bibfnamefont {H.}~\bibnamefont {Dillinger}},\ and\ \bibinfo {author} {\bibfnamefont {F.}~\bibnamefont {Pham}},\ }\bibfield  {title} {\bibinfo {title} {{Exact semiclassical expansions for one-dimensional quantum oscillators}},\ }\href {https://doi.org/10.1063/1.532206} {\bibfield  {journal} {\bibinfo  {journal} {J. Math. Phys.}\ }\textbf {\bibinfo {volume} {38}},\ \bibinfo {pages} {6126} (\bibinfo {year} {1997})}\BibitemShut {NoStop}%
\bibitem [{\citenamefont {Voros}(1983)}]{Voros:1983ret}%
  \BibitemOpen
  \bibfield  {author} {\bibinfo {author} {\bibfnamefont {A.}~\bibnamefont {Voros}},\ }\bibfield  {title} {\bibinfo {title} {The return of the quartic oscillator. the complex wkb method},\ }in\ \href@noop {} {\emph {\bibinfo {booktitle} {Annales de l'IHP Physique th{\'e}orique}}},\ Vol.~\bibinfo {volume} {39}\ (\bibinfo {year} {1983})\ pp.\ \bibinfo {pages} {211--338}\BibitemShut {NoStop}%
\bibitem [{\citenamefont {Teaney}(2006)}]{Teaney:2006nc}%
  \BibitemOpen
  \bibfield  {author} {\bibinfo {author} {\bibfnamefont {D.}~\bibnamefont {Teaney}},\ }\bibfield  {title} {\bibinfo {title} {{Finite temperature spectral densities of momentum and R-charge correlators in N=4 Yang Mills theory}},\ }\href {https://doi.org/10.1103/PhysRevD.74.045025} {\bibfield  {journal} {\bibinfo  {journal} {Phys. Rev. D}\ }\textbf {\bibinfo {volume} {74}},\ \bibinfo {pages} {045025} (\bibinfo {year} {2006})},\ \Eprint {https://arxiv.org/abs/hep-ph/0602044} {arXiv:hep-ph/0602044} \BibitemShut {NoStop}%
\end{thebibliography}

%%%%%%%%%%%%%%%%%%%%%%%%%%%%%%%%%%%%%%%%%%%%%%%

\end{document}